\pdfoutput=1
\documentclass[10pt,conference,twocolumn,final]{IEEEtran}
\usepackage[english]{babel}
\usepackage{amsmath,amsfonts,amssymb}
\usepackage{booktabs,tabularx}
\usepackage{graphicx}
\usepackage{bmpsize}
\usepackage{ifpdf}
\usepackage{enumitem}
\usepackage[usenames]{xcolor}
\usepackage{tikz}
\usepackage{textcomp}
\usepackage{hyperref}

\ifCLASSOPTIONcompsoc
    \usepackage[caption=false,font=normalsize,labelfont=sf,textfont=sf]{subfig}
\else
    \usepackage[caption=false,font=footnotesize]{subfig}
\fi
\usepackage{url}

\newcommand{\eref}[1]{(\ref{#1})}
\newcommand{\fref}[1]{Fig.~\ref{#1}} 
\newcommand{\sref}[1]{Section~\ref{#1}}
\newcommand{\tref}[1]{Table~\ref{#1}}

\graphicspath{
  {images/}
}
\ifpdf
\DeclareGraphicsExtensions{.pdf,.png}
\else
\DeclareGraphicsExtensions{.eps}
\fi

\IEEEoverridecommandlockouts

\newcommand\copyrighttext{%
	\footnotesize \textcopyright 2020 IEEE. Personal use of this material is permitted.
	Permission from IEEE must be obtained for all other uses, in any current or future
	media, including reprinting/republishing this material for advertising or promotional
	purposes, creating new collective works, for resale or redistribution to servers or
	lists, or reuse of any copyrighted component of this work in other works.
	DOI: \href{https://doi.org/10.1109/VTC2020-Spring48590.2020.9128458}{10.1109/VTC2020-Spring48590.2020.9128458}}
\newcommand\copyrightnotice{%
	\begin{tikzpicture}[remember picture,overlay]
	\node[anchor=north,xshift=0pt,yshift=-10pt] at (current page.north) {\fbox{\parbox{\dimexpr\textwidth-\fboxsep-\fboxrule\relax}{\copyrighttext}}};
	\end{tikzpicture}%
}

\title{A mmWave Bridge Concept to Solve the Cellular Outdoor-to-Indoor Challenge}

\begin{document}
\bstctlcite{ShortCTL:BSTcontrol}

\author{
\IEEEauthorblockN{
    Adrian Schumacher\textsuperscript{*}\textsuperscript{\#}, 
    Ruben Merz\textsuperscript{*} and
    Andreas Burg\textsuperscript{\#}}
\IEEEauthorblockA{\textsuperscript{*}
    Enterprise Architecture \& Innovation, Swisscom Ltd., CH-3050 Bern, Switzerland\\
    \{adrian.schumacher,ruben.merz\}@swisscom.com}
\IEEEauthorblockA{\textsuperscript{\#}
    Telecommunications Circuits Laboratory, EPFL, CH-1015 Lausanne, Switzerland\\
    andreas.burg@epfl.ch\vspace{-0.5\baselineskip}}
}
\maketitle

\copyrightnotice

\begin{abstract}
  Wireless indoor coverage and data capacity are important aspects of cellular networks.
With the ever-increasing data traffic, demand for more data capacity indoors is also growing.
The lower frequencies of the legacy frequency bands of macro outdoor cells manage to provide coverage inside buildings, however, new frequencies foreseen for the 5th generation (5G) of mobile communications in the millimeter wave (mmWave) spectrum penetrate very poorly into buildings.
Therefore, a massive densification of the network would require to deploy a large number of indoor small cells, which would lead to high deployment costs to install the necessary wired/optical backhaul.
Hence, other methods are needed that allow an increase of the data capacity indoors, bearing a lower cost than a fiber deployment.
We propose a cost-efficient out-of-band repeater architecture that provides more data capacity indoors than an outdoor macro/micro network can provide to indoor, without adversely affecting a legacy network, and which readily works with the established cellular infrastructure as well as standard handsets/smartphones.
This proposal is compared to conventional in- and out-of-band repeaters and relay nodes in order to highlight the advantages of our solution.
While the data capacity for a single link is similar to that of repeaters and relays, a macro cell can be effectively offloaded. Cell capacities corresponding to at least 3--4 times that of a repeater or relay solution can be provided, depending on the number of parallel installed links and the bandwidth in the mmWave spectrum.

\end{abstract}
\section{Introduction}
\label{sec:introduction}

Mobile data traffic is increasing globally. To keep up with the data capacity demands, new wireless communication standards are defined that can provide higher data rates by using the assigned cell bandwidth more efficiently, employing spatial multiplexing multiple input multiple output (MIMO), and more spectrum.
For the 5th generation (5G) of mobile cellular communications, new spectrum with wide bandwidths is becoming available for mobile use, which is harmonized at the World Radiocommunication Conference 2019. The spectrum in the range of 24.25\,GHz to 52.6\,GHz (and later up to 86\,GHz) is commonly referred to as millimeter wave (mmWave) spectrum. With New Radio (NR), the 3rd Generation Partnership Project (3GPP) standardized the radio access technology (RAT) for 5G, which can also operate in the mmWave spectrum.
However, statistics from studies \cite{abiresearch_inbuildingtraffic_2016} show that around 80\,\% of the mobile data traffic originates or terminates indoors. Unfortunately, mmWave frequency bands experience a much higher attenuation for penetrating into buildings compared to frequencies below 6\,GHz (sub-6\,GHz).
Still, even at 3.5\,GHz the outdoor-to-indoor coverage is challenging as reported in \cite{Schumacher_3.5GHzCoverageAssessment_2019}.
While the attenuation of certain building materials such as wood, plasterboard, and drywall does not change much with an increasing radio frequency (RF), it can significantly increase for brick and concrete walls (from 17.7\,dB at $<$3\,GHz to 175\,dB at 40\,GHz) \cite{Pi_IntroductionMillimeter-waveMobileBroadband_2011,ITUR_CompilationMeasurementDataRelating_2017}.
Therefore, even if the mmWave frequency bands can provide large data rates due to the large bandwidth, they are not suited for outdoor-to-indoor coverage.

\subsection{Small Cells}
\label{sec:intro_smallcell}
To provide high data capacity in buildings, the ideal solution is to install small cells (this term is used here as an umbrella term for various forms including picocells and femtocells).
Because these small cells (there may be multiple installed in a larger building) also need an appropriate backhauling to provide the high data rates, expensive, large scale deployment of fiber to the home (FTTH) will be required. While efforts in this direction are ongoing, it is economically not feasible to achieve this within the next couple of years and in larger countries it might take decades if not discarded completely.
Another disadvantage of the small cell may be the interference between the outdoor macro and micro cells with the indoor small cell, in case they use the same frequency bands \cite{Larson_DeploymentOptionsProvidingIndoor_2015}.
Furthermore, even reduced complexity small cells require complex hardware, which is costly and require regular maintenance and updates to follow the evolution of standards.

\subsection{Repeaters}
\label{sec:intro_repeater}
Layer~1 (amplify-and-forward) repeaters are a widely deployed inexpensive method to extend the coverage and bring data capacity to remote areas and into buildings where the outdoor macro network cannot reach.
A repeater (red box in \fref{fig:O2ISimuArch}) receives the donor base station signal through a (directional) antenna, filters the signal, and retransmits it amplified through the service antenna in the same frequency band as the donor cell.
Its operation is therefore transparent for the base stations and the user equipment (UE), which reduces hardware and integration complexity, and maintenance cost.
Considering the data capacity, two MIMO layers are standard since the introduction of Long Term Evolution (LTE), and with 5G NR even four MIMO layers may become common especially indoors. To support multiple MIMO layers, multiple donor as well as service antennas need to be installed, with the same number of repeater amplifier chains \cite{Wirth_LTEAmplifyForwardRelaying_2010}.
The major disadvantage is the additional noise that is added to the repeated signal due to the amplifier. If multiple repeater systems are deployed in a cell, they add considerable noise in the uplink of the common donor cell \cite{Hiltunen_UsingRFRepeatersImprove_2006}.
The signal delay through a repeater can be important if the power level difference between the outdoor donor cell and the indoor service cell is not sufficiently high. If the difference is small and the repeater signal delay exceeds the cyclic prefix (CP) duration of the orthogonal frequency division multiplexing (OFDM) modulation, the delayed signal from the repeater interferes with the direct signal from or to the donor base station causing inter-symbol interference (ISI) \cite{Shen_Repeater-enhancedMillimeter-waveSystemsMulti-path_2015a}.

\subsection{Relays}
\label{sec:intro_relay}
\begin{figure}
	\centering
	\includegraphics[width=\linewidth]{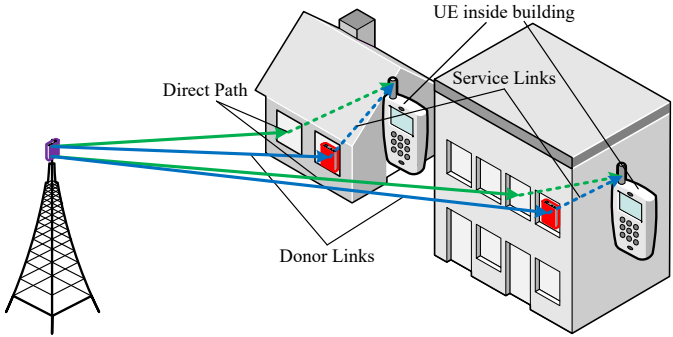}
	\caption{Base Station transmitting to mobile stations inside buildings. Direct signal path in green, relayed/repeated signal path in blue.\label{fig:O2ISimuArch}}
\end{figure}

Another method is to operate the relay on layer~2 or layer~3. A layer~2 relay decodes the received signal and sends it again on the same frequency, encoded on the transmit-side (decode-and-forward), ideally clean of noise and interference \cite{Venkatkumar_RelayingResultsIndoorCoverage_2010,Biswas_PerformanceRelayAidedMillimeter_2016}. The layer~3 relay not only decodes the received signal, but also creates a new cell with its own cell identification on either the same or on a different carrier frequency.
If the received signal level or quality is too low, the decoding may fail and the relay fails to transmit the cell signal. Therefore, a careful planning and installation of the donor antenna is required.
A disadvantage of the relay is the substantial delay from decoding and encoding the signal, which requires a high isolation between the outdoor (donor) and indoor (service) cells to prevent interference from the direct path (see \fref{fig:O2ISimuArch}). However, in mmWave spectrum, the high building entry losses help to reduce such interference.
While a layer~2 relay needs radio control functions for the communication between the donor base station and the relay, a layer~3 relay has less impact on the standard, but requires the functionality of a base station itself, much like a small cell.
Therefore, relays suffer from the same complexity and maintenance issues as small cells.

\subsection{Contribution and Outline}
In the next section we propose and describe the mmWave Bridge (mmWB), a paired out-of-band repeater architecture that employs mmWave spectrum for the signal link between the donor mmWB node installed at the base station and the service mmWB node installed at a building, but conveys legacy signals in sub-6\,GHz bands on each end.
Unlike small cells, repeaters, and relays, this simple and cost-efficient solution provides indoor capacity without impacting the legacy macro network with additional traffic or interference, and without costly wired (optical fiber) backhaul to every building. Furthermore, it allows scaling up to cover multiple buildings and floors subject to the available bandwidth in the mmWave spectrum and the feasible mmWave antenna beamwidth.
\sref{sec:sysmodels} describes the link budget models for the macro outdoor-to-indoor, small cell, in-band repeater, layer-3 relay, and mmWave bridge. These methods are compared using simulations and the results are presented and discussed in \sref{sec:results}.
Finally, we conclude with a summary of the identified advantages.

\section{mmWave Bridge}
\label{sec:mmwb}

With the term mmWave Bridge we refer to a system of two paired nodes, where the first node translates a sub-6\,GHz signal to mmWave spectrum and transmits this signal to the second node, which translates the signal back to the original frequency.
Similar concepts to our mmWave bridge have been proposed earlier.
In \cite{MBreiling_UE-SideVirtualMIMOUsing_2014} low-cost amplify-and-forward relay nodes are introduced that translate between spatially multiplexed MIMO donor signals in sub-3\,GHz bands and frequency multiplexed service signals in mmWave. By deploying them in great numbers indoors, large capacity gains because of higher MIMO order and multi-user MIMO are claimed. The disadvantage is that the base station as well as the UE need to incorporate special RAT protocol stacks (either based on IEEE~802.11ad or special functions for this type of relay which are not yet part of 5G NR). Furthermore, it is envisaged that the nodes shall be low-cost and intended for customer self-installation. Without high-quality hardware and without installation planning and guidance, this approach may not always provide the expected high capacity.
A similar approach and closer to our mmWave bridge proposal is presented in \cite{Zhu_Millimeter-waveMicrowaveMIMORelays_2018}. Instead of relying on sub-3\,GHz signals to penetrate into buildings, an outdoor unit uses mmWave frequencies for the donor link and the unlicensed 2.4\,GHz band for the service link. To achieve high data capacity, multiple spatial MIMO layers are frequency multiplexed in the mmWave frequency domain on the donor link and translated to the common frequency for the service link. The disadvantages are that a time-space encoder and decoder are required as an integral part inside the base stations serving the relays, and that the 2.4\,GHz band is often crowded because of the many Wi-Fi access points, Bluetooth devices, and other applications using this unlicensed band.

\begin{figure}
	\centering
	\includegraphics[width=\linewidth]{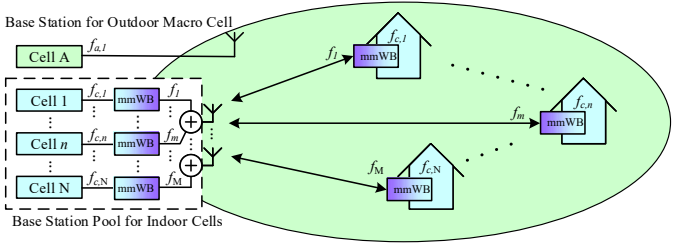}
	\caption{Multiple indoor cells (blue) fronthauled over mmWave Bridges (boxes labeled mmWB; donor bridge nodes on the left, service bridge nodes at the houses on the right) in the coverage area of a macro cell (green large circle)\label{fig:principle_mmwave_bridge}}
	\vspace{-0.5em}
\end{figure}

As explained above, the mmWave bridge consists of two complementing nodes, shown in \fref{fig:principle_mmwave_bridge} for a deployment example. The donor bridge node is installed at the base station pooling location. It upconverts the downlink (DL) cell signals from the sub-6\,GHz frequency, designated with $f_{c,n}$, where $n$ denotes the cell $[1,\text{N}]$, to the mmWave frequencies $f_m$, where $m = [1,\text{M}]$, for transmission over the donor link.
Using high gain (analog) beamforming antennas, the cells are transmitted to the service bridge node which is known as the customer premise equipment (CPE). Note that the large bandwidth (several 100\,MHz) in mmWave spectrum allows for frequency-multiplexing of many sub-6\,GHz cells with bandwidths of 20-100\,MHz. Thanks to the directionality of the high gain beamforming antennas, the mmWave carrier frequencies can also be reused depending on the angular spread, thus $\text{M} \leq \text{N}$.
The CPE consists of an outdoor and indoor unit. The outdoor unit is installed on a window or the building wall ideally with line-of-sight (LOS) to the antenna of the donor bridge. Depending on the size of the building and the required data capacity, one CPE may be sufficient, or multiple CPEs are installed to cover multiple floors or building sections.
The indoor unit of the CPE reradiates the downconverted DL cell signal on its original sub-6\,GHz frequency.
For the uplink (UL), the CPE picks up the UL signal, upconverts it to the corresponding mmWave frequency, transmits it to the donor bridge, where the signal is downconverted and fed to the base station.

\begin{figure}
	\centering
	\includegraphics[width=\linewidth]{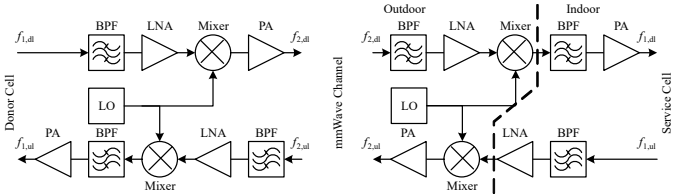}
	\caption{Block diagram of the mmWave Bridge with the donor bridge node on the left side and the service bridge node (CPE) on the right side\label{fig:schematic_mmwave_bridge}}
	\vspace{-0.5em}
\end{figure}

The high level architecture of the mmWave bridge is shown in \fref{fig:schematic_mmwave_bridge}. Note that the dashed line at the CPE on the right side only shows an exemplary separation between outdoor and indoor unit.
Even though only one signal path is shown, the default setup supports $2\times2$ MIMO on the sub-6\,GHz as well as the mmWave frequency links using cross-polarized antennas. Higher order MIMO may be supported as well.

The proposed mmWave bridge concept has many operational advantages over the existing methods. First and foremost, it does not require any changes in the existing and well operating outdoor macro networks nor does it have any adverse impact.
Second, it readily works with standard base stations and UEs already in use by customers.
Third, its simple up-/downconversion can be fully implemented in analog circuits, minimizing the signal delays.
Fourth, its transparent amplify-and-forward functionality is independent of the RAT and is forward-compatible with evolutions in LTE, NR, or future technologies (6G).
Fifth, the deployment of such a mmWave bridge is cheaper than deploying optical fiber and small cells or additional sub-6\,GHz cells.
If the CPE is carefully engineered and simple-to-follow instructions are given, it allows self-installation by customers.
Further applications are possible e.g., serving a distributed antenna system (DAS) installed along railway tracks for an RF corridor as described in \cite{Jamaly_DeliveringGigabitCapacitiesPassenger_2019}. In this case, a donor bridge node would serve multiple service bridge nodes installed in regular distances e.g., on catenary masts. In another application, the mmWave bridge could serve front-haul links from a site with optical fiber access to multiple locations where it is challenging to bring optical fiber e.g., in extremely hilly terrain.

\section{System Models}
\label{sec:sysmodels}
The performance of our proposed mmWave bridge is evaluated against the competing methods discussed in \sref{sec:introduction} and a macro outdoor network as a baseline.
For the sake of brevity, we only consider a 5G NR carrier at 3.6\,GHz with 100\,MHz of bandwidth and one at 26\,GHz with 400\,MHz of bandwidth. Note that 26\,GHz is a 5G mmWave candidate band for Europe.

The scenario consists of $N_\text{B}$ individual buildings scattered across a circular sector of 120$^\circ$ at uniformly distributed distances\footnote{Analysis of building location statistics around existing base stations show a close to uniform distribution, i.e., the farther away from a base station, the sparser the building density} from the base station, a 5G Node-B (gNB). The cell radius with the considered frequencies and transmit powers is limited to several hundreds of meters, over which the terrain usually does not change much, therefore a flat terrain is assumed.
For simplicity, buildings occupy a square area, are all facing the gNB, and all have the same height.
If an outdoor donor antenna at a specific height above ground on the building wall (or window) is shadowed by another building, non line of sight (NLOS) condition is assumed, otherwise line of sight (LOS). The 3GPP empirical path loss models \cite{zz3gpp.38.901} are used to compute the outdoor propagation loss $L_\text{out}$ up to the building wall.
For methods with an indoor service antenna, LOS is assumed for a uniformly distributed distance $d$ between 2\,m and 10\,m (e.g., in a living room, office, shop), and the free-space path loss model (based on the Friis transmission equation) is used to model the indoor path loss $L_\text{in}$.

Simulation parameters are summarized in \tref{tab:simu_params}. Note that transmit powers are specified as effective isotropic radiated power (EIRP) and therefore include the antenna gain and feeder losses.
Equations are given for the DL direction. The UL experiences the same path losses and the equations can be derived analogously.
Noise is neglected in the following equations, if it is orders of magnitude below the thermal noise floor, however, the noise is accurately computed in the simulations.

Interference from neighboring cells has not been considered, which would mainly degrade the performance of the macro outdoor-to-indoor, layer~1 repeater, and layer~3 relay methods in the sub-6\,GHz spectrum. In the mmWave spectrum, neighbor cell interference is lower thanks to beamforming and the higher path loss.
Due to the static links mostly with LOS condition and the wide bandwidth used, fading effects are neglected.

\subsection{Models for Method Evaluation}
\subsubsection{Outdoor Macro Network}
The first method serves as a baseline with a macro outdoor network serving indoor users. The building entry loss (BEL) from outdoor-to-indoor is computed according to the model ITU-R P.2109 \cite{ITUR_PredictionBuildingEntryLoss_2017} and denoted with $L_\text{O2I}$. This model allows the calculation of a BEL for frequencies of 80\,MHz up to 100\,GHz. The median probability is used for `traditional' buildings i.e., without thermally efficient and highly attenuating building materials such as metallic coated glass.
Note that this BEL model already accounts for the indoor propagation loss and atmospheric gas attenuation, therefore the received power for the UE becomes $P^\text{MN}_\text{UE,r} = (P_\text{BS,t} G_\text{UE}) / (L_\text{out} L_\text{O2I})$, with $P_\text{BS,t}$ denoting the constant base station transmit power and $G_\text{UE}$ the UE antenna gain.

\subsubsection{Small Cell Model}
The small cell represents the ideal case where each building provides optical fiber connectivity and is equipped with a small cell. The received power for a UE becomes $P^\text{SC}_\text{UE,r} = (P_\text{SC,t} G_\text{UE}) / L_\text{in}$, with $P_\text{SC,t}$ denoting the constant small cell transmit power.
The only performance impact for this method is the interference from the outdoor macro cell that is assumed to operate also on the same frequency as the indoor small cell. Therefore, the interference power becomes $P^\text{SC}_\text{UE,if} = P^\text{MN}_\text{UE,r}$.

\subsubsection{Repeater Model}
The layer~1 repeater method is modeled as an in-band amplify-and-forward relay. It is assumed that one repeater is installed at each building, serving the indoor users with an amplified signal from outside the same building.
The repeater model consists of an amplifier with a limited maximum output power $P_\text{R,t,max}$, additive noise, and a signal delay that can cause ISI as explained in \sref{sec:intro_repeater}.
In time domain, the signal $y(t)$ at the repeater output is calculated from the outside signal $x(t)$ at the building, according to \eref{eq:repeater_TD}
\begin{equation}
\label{eq:repeater_TD}
y(t) = \frac{G_\text{R}}{L_\text{in}} \sum\limits_{k=1}^{\infty} \left(\frac{G_\text{R}}{L_\text{iso}}\right)^{k-1} x(t - k \tau) \quad\text{,}
\end{equation}
where $G_\text{R}$ denotes the repeater gain including all antenna gains, $L_\text{iso}$ the isolation loss between the outdoor donor antenna and the indoor service antenna of the repeater, and $\tau$ is the repeater signal delay. Repeater self-interference is added to the system for $k>1$.
With $P_\text{R,UE}$ denoting the power of the signal $y(t)$ from the repeater at the UE antenna (excluding the UE antenna gain $G_\text{UE}$) and $P_\text{R,UE}L_\text{in} \leq P_\text{R,t,max}$, and a reasonably large ratio of $G_\text{R}/L_\text{iso}$ (e.g., $>$15\,dB), the received power for the UE is approximated as:
\begin{equation}\label{eq:repeater_P}
P^\text{L1R}_\text{UE,r} = \left\{
	\begin{array}{lr}
		(P_\text{R,UE} + P^\text{MN}_\text{UE,r}) G_\text{UE} & : \tau<T_\text{CP}\\
		P_\text{R,UE} G_\text{UE} & : \tau \ge T_\text{CP}
	\end{array}\right.
\end{equation}
If the repeater signal delay $\tau$ is equal or greater than the length of the CP, the direct signal penetrating from outside to inside is added to the interference power $P^\text{L1R}_\text{UE,if}$, which also sums the power from the repeater self-interference (i.e., the power of $y(t)$ for $k>1$), all multiplied by $G_\text{UE}$, the antenna gain of the UE.
The output noise power is computed according to $P^\text{L1R}_\text{N,t} = N_\text{0} F_\text{L1R}(G_\text{R}) B G_\text{L1R,S}$, where $N_\text{0}$ denotes the thermal noise power spectral density (-174\,dBm/Hz), $F_\text{L1R}$ denotes the noise factor as a function of the repeater gain, $B$ is the signal bandwidth, and $G_\text{L1R,S}$ is the service antenna gain.

\subsubsection{Layer~3 Relay Model}
We only consider a layer~3 relay because it offers the out-of-band operation that a layer~2 relay does not support (on the link budget level considered here, a layer~2 relay performs very similar to a layer~3 relay). Also here, it is assumed that one relay is installed at each building with an outdoor donor antenna facing the gNB and (an) indoor service antenna(s).
The relay model consists of a low noise amplifier with a specific noise figure, and a constant output signal power.
The equations for the power computation are analog to those for the small cell, with $P^\text{L3R}_\text{UE,r} = (P_\text{L3R,t} G_\text{UE}) / L_\text{in}$, and $P_\text{L3R,t}$ denoting the layer~3 relay transmit power.
The signal processing delay is orders of magnitude larger than the CP duration $T_\text{CP}$ of 5G NR, therefore an outdoor macro cell interferes with the indoor cell of the relay, depending on the BEL and if they both use the same frequency. In that case the interference becomes $P^\text{L3R}_\text{UE,if} = P^\text{MN}_\text{UE,r}$.

\subsubsection{mmWave Bridge Model}
Finally, for the mmWave bridge we also assume that one CPE is installed at each building and served from the gNB location.
Each amplifier in the donor bridge node as well as the CPE contributes to the noise in the overall link budget. Similar to the repeater, the amplifier gain is fixed but limited by the maximum specified output power $P_\text{mmWB,t,max}$. A signal delay is also added, but thanks to the out-of-band operation, there is no self-interference in the system.
The received power at the UE becomes
$P^\text{mmWB}_\text{UE,r} = (G_\text{mmWB,D} P_\text{mmWB,t} G_\text{UE}) / (L_\text{out} L_\text{in})$, where $G_\text{mmWB,D}$ represents the CPE donor antenna gain and $P_\text{mmWB,t} = P_\text{mmWB,r} G_\text{mmWB} \leq P_\text{mmWB,t,max}$ denotes the CPE indoor transmit power.
As in all other methods, interference from an outdoor macro cell is considered on the same frequency as the indoor cell: $P^\text{mmWB}_\text{UE,if} = P^\text{MN}_\text{UE,r}$.
The output noise power is computed according to $P^\text{mmWB}_\text{N,t} = N_\text{0} F_\text{mmWB}(G_\text{mmWB}) B G_\text{mmWB,S}$, where $F_\text{mmWB}$ denotes the CPEs noise factor as a function of the mmWave bridge gain, and $G_\text{mmWB,S}$ denotes the service antenna gain.

\begin{table}
	\centering
	\caption{Simulation parameters and values}
	\label{tab:simu_params}
	\vspace{-0.8em}
	\begin{tabularx}{\linewidth}{@{}lXX@{}}
		\toprule
		Parameter & \multicolumn{2}{l}{Values} \\
		\midrule
		Carrier frequency $f_\text{c}$ & \textbf{3.6\,GHz} & \textbf{26\,GHz} \\
		Carrier bandwidth $B$ & 100\,MHz & 400\,MHz \\
		Subcarrier spacing & 30\,kHz & 120\,kHz \\
		CP duration $T_\text{CP}$ & 2.3\,$\mu$s & 0.6\,$\mu$s \\
		TDD DL/UL ratio & 4:1 & 4:1 \\
		gNB $P_\text{tx,EIRP}$ & 656\,W & 1000\,W \\
		gNB UL noise figure & 3\,dB & 9\,dB \\
		gNB antenna gain & 24\,dBi & 27\,dBi \\
		gNB $P_{0\text{,nom,PUSCH}}$ & -105\,dBm & -105\,dBm \\
		UE max. TX power & 23\,dBm & 23\,dBm \\
		UE DL noise figure & 8\,dB & 9\,dB \\
		UE antenna gain & 3\,dBi & 9\,dBi \\
		\midrule
		Outdoor propagation & \multicolumn{2}{l}{path loss according to 3GPP RMa LOS/NLOS} \\
		\quad parameters values & \multicolumn{2}{l}{$h_{\text{BS}} = 30\,\text{m}$, $h_{\text{UT}} = 2\,\text{m}$, $W = 20\,\text{m}$, $h = 10\,\text{m}$} \\
		\midrule
		Building entry loss & \multicolumn{2}{l}{according to ITU-R P.2109-0} \\
		\quad parameters values & \multicolumn{2}{l}{$P = 0.5$, traditional, $\theta = 0$} \\
		\midrule
		Small cell & \multicolumn{2}{l}{$P_\text{max,EIRP}$ = 30\,dBm, $P_{0\text{,nom,PUSCH}}$ = -105\,dBm} \\
		\quad noise figure & 7\,dB & 9\,dB \\
		\quad antenna gain & 6\,dBi & 24\,dBi \\
		\midrule
		L1 repeater & \multicolumn{2}{l}{max. DL \& UL gain = 60\,dB, delay = 1\,$\mu$s} \\
		\quad donor side & \multicolumn{2}{l}{$P_\text{max,EIRP}$ = 40\,dBm, noise figure = 9\,dB} \\
		\quad service side & \multicolumn{2}{l}{$P_\text{max,EIRP}$ = 30\,dBm, noise figure = 9\,dB} \\
		\quad donor antenna gain & 6\,dBi & 24\,dBi \\
		\quad service antenna gain & 6\,dBi & 12\,dBi \\
		\midrule
		L3 relay & \multicolumn{2}{l}{RX sensitivity -125\,dBm (RSRP), delay = 10\,ms} \\
		\quad donor side & \multicolumn{2}{l}{$P_\text{max,EIRP}$ = 40\,dBm, noise figure = 6\,dB} \\
		\quad service side & \multicolumn{2}{l}{$P_\text{max,EIRP}$ = 30\,dBm, noise figure = 6\,dB} \\
		\quad donor antenna gain & 6\,dBi & 24\,dBi \\
		\quad service antenna gain & 6\,dBi & 12\,dBi \\
		\midrule
		mmWave Bridge & \multicolumn{2}{l}{max. DL \& UL gain = 60\,dB, delay = 50\,$\mu$s} \\
		\quad $P_\text{max,TX,EIRP}$ & \multicolumn{2}{l}{outdoor: 40\,dBm, indoor: 30\,dBm} \\
		\quad noise figure & 7\,dB & 9\,dB \\
		\quad antenna gain & 6\,dBi & 24\,dBi \\
		\bottomrule
	\end{tabularx}
	\vspace{-1em}
\end{table}

\subsection{Capacity Estimation}
\label{sec:capacity}
For each method and simulated building, the signal to interference and noise ratio (SINR) can be computed as:
\begin{equation}
\label{eq:sinr}
\gamma = \frac{P^\cdot_\text{UE,r}}{N_\text{0} \cdot B \cdot F_\text{UE} + P^\cdot_\text{N} + P^\cdot_\text{UE,if}} \quad\text{,}
\end{equation}
with $F_\text{UE}$ representing the noise factor of the UE, $P^\cdot_\text{N} = P^\cdot_\text{N,t} L_\text{in} G_\text{UE}$ the noise power received by the UE, and $P^\cdot_\text{UE,if}$ the interference power received by the UE.
The SINR denoted with $\gamma$ is then mapped to a single input single output (SISO) capacity value $C$ in bit/s using the modified Shannon-Hartley theorem according to the 5G NR technical report \cite[Sec.~5.2.7]{zz3gpp.38.803}: $C = \alpha B \log_2 (1 + \gamma)$. $B$ represents the channel bandwidth in Hertz and $\alpha$ is an attenuation factor representing implementation losses (0.6 for DL, 0.4 for UL). Also, according to \cite{zz3gpp.38.803}, the capacity $C$ is set to zero for SINR~$<$~-10\,dB. For the upper capacity limit, the equation in \cite[Sec.~4.1.2]{zz3gpp.38.306} is used to compute the supported maximum data rate.
For time division duplex (TDD) operation, one needs to also adjust for the DL to UL ratio, see \tref{tab:simu_params}.
Note that the obtained capacity is valid for additive white Gaussian noise (AWGN) channels.

\section{Results and Discussion}
\label{sec:results}

\subsection{Simulation Results}
\begin{figure}
	\centering
	\includegraphics[width=\linewidth]{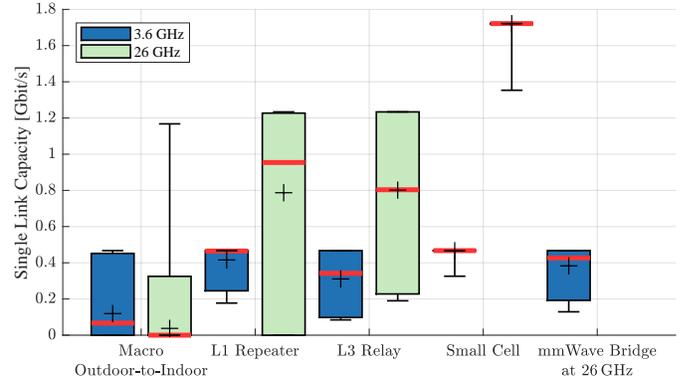}
	\caption{Single link SISO capacity comparison for the five discussed methods in two frequency bands, where applicable. The boxes show the 5\%ile and 95\%ile, the median (horizontal red line inside the box), and the mean value (+ marker)\label{fig:linkcapacomp}}
\end{figure}

Monte Carlo simulations are performed with $N_\text{MC}$ = 100 iterations for the distribution of $N_\text{B}$ = 300 buildings. The computed capacities for a single SISO link are shown in \fref{fig:linkcapacomp} with the boxes showing the 5\%ile and 95\%ile, the horizontal green lines inside the boxes the median, and the plus markers the mean values.

\begin{figure}
	\centering
	\includegraphics[width=\linewidth]{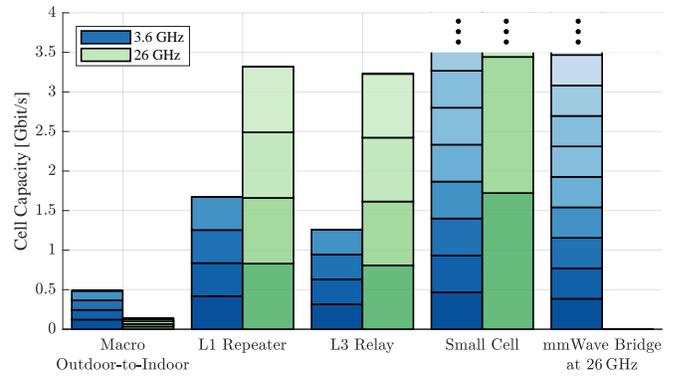}
	\caption{Cell capacity comparison based on simultaneously scheduled single SISO links (bar segments in different color shades) for the five discussed methods in two frequency bands, where applicable\label{fig:cellcapacomp}}
\end{figure}

The 5G NR standard supports multi-user MIMO (MU-MIMO) with up to four simultaneously scheduled UE. In a full-buffer scenario, we can assume an upper bound of four simultaneous links for the outdoor-to-indoor macro baseline, the layer~1 repeater, and the layer~3 relay. This is also shown in \fref{fig:cellcapacomp} with the stacked bars representing co-scheduled users based on the mean link capacity. Note that the small cell method ideally scales the total capacity in the area with the number of installed small cells in the buildings. The mmWave bridge however, only scales depending on the number of available mmWave bridge links.

\subsection{Discussion}
The mean SISO capacity available indoors from an outdoor macro network is rather low for 3.6\,GHz (119\,Mbit/s) and close to zero for mmWave, see the left two boxes in \fref{fig:linkcapacomp}.
Adding a layer~1 repeater generally improves the capacity (416\,Mbit/s for 3.6\,GHz, 787\,Mbit/s for 26\,GHz), but it is crucial that the repeater delay is smaller than the CP length to prevent ISI.
The layer~3 relay does not perform as well as a layer~1 repeater for buildings closer to the base station due to the inter-cell interference. However, for higher path losses especially in mmWave, the layer~3 relay outperforms the layer~1 repeater. The mean SISO capacity with a layer~3 relay reaches 311\,Mbit/s for 3.6\,GHz and 801\,Mbit/s for 26\,GHz.
An optimally placed small cell manages to deliver consistently high capacities for a user in the same room (467\,Mbit/s for 3.6\,GHz, 1.721\,Gbit/s for 26\,GHz). The only capacity impacting factor is inter-cell interference from the outdoor macro cell.
A mmWave bridge operating at 26\,GHz can provide almost as much capacity as a layer~1 repeater (383\,Mbit/s for the 3.6\,GHz cell). The difference comes from the fact that the mmWave bridge consists of more amplifiers and components that add more noise to the signal than a layer~1 repeater.

Buildings are increasingly built or equipped with metallic coated Low-E windows. If the ITU-R P.2109 model for `modern' buildings instead of `traditional' is used for the BEL, only the macro outdoor-to-indoor capacity results are affected: at mmWave frequencies, hardly any communication from inside a building is possible with an outdoor mmWave small- or macro cell, thus outdoor mmWave cells would not cover indoor UE.

The large single user capacity gains that the system proposed in \cite{MBreiling_UE-SideVirtualMIMOUsing_2014} might provide, cannot be reached with the mmWave bridge.
However, the significant advantage of the mmWave bridge is that no change in the deployed networks and at UE are required. Another advantage comes from multiplexing multiple links in the spatial and frequency domain. Assuming a mmWave antenna gain of 24\,dBi, the half power beamwidth results to around 10$^\circ$. For a cell sector with 120$^\circ$ in azimuth and minimal angular spread, up to 12 antennas can be used to cover the sector, scaling the capacity by a factor of 12, resulting in a 2.8-fold increase compared to the layer~1 repeater cell capacity. If four 100\,MHz cells are frequency multiplexed in the 400\,MHz mmWave bandwidth, the scaling factor even becomes 48, yielding an 11-fold increase compared to the layer~1 repeater cell capacity. Referring to \fref{fig:cellcapacomp}, we can note that minimum 5 mmWave bridge links provide already more total cell area capacity than a single 5G NR gNB at 3.6\,GHz. Furthermore, minimum 9 mmWave bridge links outperform the available cell capacity when using repeaters or relays in the 26\,GHz mmWave band.

\section{Conclusion}
\label{sec:conclusions}

With 5G, much higher frequencies can be used for mobile communications than before. The higher frequencies allow for much larger signal bandwidth and data capacity. However, they experience higher path losses and penetrate much less into buildings. To solve this challenge, several methods have been reviewed: small cell, layer~1 in-band repeater, layer~3 in- and out-of-band relay, and a mmWave bridge has been proposed as a simpler concept.
The mmWave bridge frequency-multiplexes sub-6\,GHz cells in mmWave spectrum for the outdoor donor link from the base station to the buildings, but re-radiates the sub-6\,GHz cells on their original carrier frequencies inside the buildings.
This brings many advantages: no changes are necessary in existing networks nor are there any adverse impacts when deploying mmWave bridges; the concept readily works with standard base stations and available UEs; the simple architecture can be fully implemented in analog circuits, minimizing the signal delays; its transparent amplify-and-forward architecture works independent of the RAT and is forward-compatible.
Simulations have been performed under given assumptions. They show that a single mmWave bridge performs almost as good as an in-band layer~1 repeater, but the high spatial- and frequency-multiplexing gains allow for a many-fold (2.8 to 11 in the studied example) cell capacity increase, compared to a layer~1 repeater or layer~3 relay deployment based on a 5G NR cell.

\bibliographystyle{IEEEtran}
\bibliography{references}

\end{document}